\documentstyle[aps, 12pt,epsf]{revtex}
\input psfig

\begin{document}

\title{Quasicrystals in a Monodisperse System}

\author{Anna Skibinsky$^{1,2}$, Sergey V. Buldyrev$^1$, Antonio Scala$^1$,\\
        Shlomo Havlin$^{1,3}$, H. Eugene Stanley$^1$}

\address{$^1$Center for Polymer Studies and Department of Physics,\\
          Boston University, Boston, MA 02215, USA\\
         $^2$Department of Chemistry,
          Boston University, Boston, MA 02215, USA\\
         $^3$Department of Physics,
          Bar-Ilan University, Ramat-Gan 52900, Israel }

\maketitle

\vspace{5pt}

\begin{abstract}
We investigate the formation of a two-dimensional quasicrystal in a
monodisperse system, using molecular dynamics simulations of hard sphere
particles interacting via a two-dimensional square-well potential. We
find that more than one stable crystalline phase can form for certain
values of the square-well parameters.  Quenching the liquid phase at a
very low temperature, we obtain an amorphous phase.  By heating this
amorphous phase, we obtain a quasicrystalline structure with five-fold
symmetry.  From estimations of the Helmholtz potentials of the stable
crystalline phases and of the quasicrystal, we conclude that the
observed quasicrystal phase can be the stable phase in a specific range
of temperatures.
\end{abstract} 


\vspace{30pt}
\section{Introduction}

Stable quasicrystalline phases are typically found in binary mixtures
\cite{SCH}, where the various arrangements of the two components
contribute to the degeneracy of the local environments \cite{WIDOM},
allowing a quasicrystalline phase to be entropy stabilized \cite{WOJ}.
With one notable exception \cite{JAGLA}, previous studies
did not support the existence of a stable quasicrystalline phase in a
monodisperse system interacting with a simple potential \cite{SUBN,APS}.

We study a simple model that allows us to estimate the crystal and
quasicrystal entropies and thereby study the Helmholtz potentials of the
crystals and quasicrystal.  The ground state of this system is a
periodic crystal, yet we explore the possibility that the
quasicrystalline configuration is the equilibrium state in a certain
temperature regime.  Although quasicrystals do not have short range
order, they do have recurring local environments that, in our model,
resemble the basic cells of the stable crystalline phases. From the
entropies of the stable crystalline phases and by estimating the
configurational entropy of the quasicrystal, we infer that the
quasicrystal may be an equilibrium state. We observe sharpening of five
fold diffraction peaks when the starting amorphous phase is annealed. In
two dimensions, five fold diffraction peaks pertain to
crystallographically disallowed point groups which characterize
quasicrystals \cite{LUBE}.

\section{MD Methods}

To study quasicrystalline stability in a monodisperse system, we perform
molecular dynamics (MD) simulations of a two-dimensional model of hard
spheres interacting with an attractive square-well (SW) potential
[Fig.~\ref{sw_pot}].  The simplicity of this SW potential allows us to
study the fundamental characteristics of the system. By tuning the width
of the SW potential, we can control the local geometric configurations
formed by the particles.  The structures of the crystalline and
quasicrystalline phases can thus be clearly defined and analyzed.

We perform MD simulations in the NVT ensemble, using a standard
collision event list algorithm \cite{RAPAPORT} to evolve the system,
while we use a method similar to the Berendsen method to achieve the
desired temperature \cite{SERGMETH}.  The depth of the potential well is
$\epsilon=-1.0$.  Energies are measured in units of $\epsilon$,
temperature is measured in units of energy divided by the Boltzmann
constant, $\epsilon/k_{B}$, and the mass of the particle is $m=1$. We
choose the value of the hard core distance to be, $a=10$, and the ratio
of the attractive distance $b$ to the hard core distance
$a$, to be $b/a=\sqrt{3}$.  Since the diagonal distance between two
corners of a square is $\sqrt{2}$ times the length of one side, choosing
$b/a=\sqrt{3}$ favors the formation of a square crystal lattice where
each particle interacts with 8 neighbors [Fig.~\ref{3xtals}a]. This
constraint inhibits the formation of a triangular crystal, which would
form at low temperatures if $b/a>\sqrt{3}$ or at high densities.

\section{Crystal and Amorphous Phases}

Studying the behavior of the system at low temperatures we observe the
formation of local structures similar to that shown in
Fig.~\ref{3xtals}.  These structures constitute local environments
\cite{WIDOM} that can reproduce crystallographically allowed symmetry if
translationally ordered.  First, we consider the stable periodic crystal
phases produced by translationally ordering each of the
configurations in Fig.~\ref{3xtals} and calculate the energies of these
crystal structures at $T=0$.  In our system, the two allowed local
configurations are the 4-particle square and the 5-particle pentagon
(indicated by the symbol ``P'' in Fig.~\ref{3xtals}b,c). Particles form
these two geometries because the nearest neighbor diagonal and
adjacent distance between particles in these configurations is less than
$b/a=\sqrt{3}$, the SW width. Four particle squares make up the square
crystal; since each particle has $8$ neighbors, at $T=0$, the potential
energy per particle is $U_{sq}=-4.0$. Pentagons do not tile the plane;
however, the formation of two kinds of crystals based on the local five
particle pentagon is possible.  In the type I pentagonal crystal, each
crystalline cell consists of $5$ particles, one of which has $8$
neighbors and four of which have $9$ neighbors; hence, $U_{pI}= -4
\frac{2}{5}$ [Fig.~\ref{3xtals}b].  In the type II pentagonal crystal,
each crystalline cell consists of $6$ particles, two of which have $8$
neighbors and $4$ of which have $9$ neighbors; hence,
$U_{pII}=-4\frac{1}{3}$ [Fig.~\ref{3xtals}c].  Since $U_{pI} < U_{pII} <
U_{sq}$, at our chosen density and low enough temperatures, the type I
pentagonal crystal should be the stable phase at $T=0$ [Fig.~\ref{fe}].

Next, we investigate the stability of the three crystalline phases at
$T>0$ by estimating the Helmholtz potential per particle $A=U-TS$ in the
square crystal and in the pentagonal crystals of type I and type
II. Here $S$ is the entropy.  Since our simulations are performed at
constant density, we must use the Helmholtz potential instead of the
Gibbs potential. We study the system at dimensionless number density
$\rho= a^{2}N/V\equiv0.857$. We have simulated a square crystal with
N=961, a pentagonal crystal type I with N=1040, and a pentagonal crystal
type II with N=792, all at the same $\rho$. We checked that at low
temperatures, $T < 0.1$, the potential energy $U(T)$ is temperature
independent, and has the same value as the potential energy of the ideal
crystal at $T=0$. Hence, we approximate $U(T)$ at higher $T$ by $U(0)$.

In order to plot the behavior of the Helmholtz potentials of the three
crystals for $T<0$, we find the entropic contributions $S$, by
estimating the entropy per particle for each of the three crystal
types. We use the probability density $p(x,y)$ to find a particle at
position $x,y$, where the average is taken over every particle in the
crystalline cell:

\begin{equation}
{S}=\bigg{\langle} \int{p(x,y) \ln p(x,y) dx dy
\bigg{\rangle}}_{cell} .
\end{equation}

\noindent
We estimate $p(x,y)$ by the fraction of the total time $t$ spent
by a particle in a discretized area, $\Delta x \Delta y$, at a low
enough temperature that the potential energy fluctuations of the
crystalline structure are negligible. The values of the entropies for the
three crystals are given in Table I.

Our estimates for the temperature dependence of the Helmholtz potential
for the three types of crystals are given in Fig.~\ref{fe}.  The
condition for stability of the pentagonal crystals is that their
Helmholtz potentials, $A_{pI}$ and $A_{pII}$, are lower than the
Helmholtz potential of the square crystal, $A_{sq}$.  In accord with
this condition, the square crystal is stable at temperatures above
$T=0.203$, the type II pentagonal crystal is stable between $T=0.195$
and $T=0.203$ and the type I pentagonal crystal is stable below
$T=0.195$.

While studying the interesting region around $T\approx0.2$
(see Fig~\ref{fe}), we observe the formation of the quasicrystal. We
choose to investigate, using MD simulations, our system at
$T\approx0.2$ because this is the temperature regime
where the three crystals have similar values of Helmholtz potential.
Cooling the fluid phase, we find the formation of the square crystal
below $T\approx0.5$. However, when further cooled into the
temperature regime where the Helmholtz potentials of the two pentagonal
crystals are lower than the Helmholtz potential of the square crystal,
the system does not form pentagonal crystal I or pentagonal crystal II
(within our simulation times), but remains as the square crystal. Hence,
we use a different approach to try to form the pentagonal crystals: we
heat an amorphous phase. We first form the amorphous phase by quenching
the system from high to very low temperatures $T\leq0.1$.
To do this, we study a system of $N=961$ \cite{FN529} particles at
$\rho=0.857$ which is initially in the fluid phase at high temperature,
$T=10$.  We quench this system to $T=0.1$
and thermalize for $10^7$ time units \cite{TIME}. Time constraints
prevent us from studying systems with more than 961 particles. Long
thermalization times are required to stabilize thermodynamic observables
like energy and pressure.

The amorphous phase is a homogeneous mixture of pentagons and squares
[Fig.~\ref{diff-fig}a]. The lack of long range structural order in the
amorphous phase is evident from the homogeneity of the circles in the
isointensity plot [Fig.~\ref{diff-fig}b]. When heating the amorphous
phase\cite{qrq} to temperatures above $T\approx0.15$, we
find that diffusion becomes sufficient for local rearrangement to occur,
and the pentagons begin to coalesce.  Instead of forming type I or II
pentagonal crystals, the pentagons begin to form rows
[Fig.~\ref{diff-fig}c] that bend at angles which are multiples of
36$^{o}$.  The angle in the bending of the rows gives rise to the
five-fold orientational symmetry, which corresponds to the ten easily
observed peaks in the isointensity plot [Fig.~\ref{diff-fig}d]. These 10
peaks are characteristic of the quasicrystal phase \cite{TSAI}, as they
are arranged with disallowed fifth order point group symmetry
\cite{LUBE}. The configuration that we obtain has defects, mainly patches
of square crystal, which cause the discontinuity in the rows and lead to
the broadening of the diffraction peaks.  For comparison, we present in
Fig.~\ref{pxtals-fig} the isointensity plots of the simulated square and
pentagonal crystals. The diffraction patterns illustrate the symmetry of
the original crystal system. The four equal sides of the square crystal
unit cell [Fig~\ref{3xtals}a] are clear in the symmetry of the
isointensity plot Fig.~\ref{pxtals-fig}. In the isointensity plot of
pentagonal crystal I [Fig.~\ref{pxtals-fig}b], the central region which
corresponds to the long range order, shows no hints of anything but well
defined centered-rectangular symmetry [Fig~\ref{3xtals}b]
\cite{KITT}. The isointensity plot of pentagonal crystal II has mainly a
rectangular symmetry that matches the rectangular symmetry of the unit
cells [Fig~\ref{3xtals}c].  Although the two pentagonal crystals are
formed from ordered pentagons, their long range symmetries are four
sided.  Their corresponding isointensity plots illustrate these four
fold symmetries which are distinctly different from the five fold
quasicrystal isointensity plot.

\section{Quasicrystal}
\subsection{Formation}

Since the phase transition between the two pentagonal crystals occurs at
$T\approx0.2$, we choose this temperature as the one to investigate for
quasicrystal formation. After the amorphous phase is quenched to T=0.1,
we anneal the system at $T=0.205$, for $2\times 10^{7}$ time units, and
calculate the diffusion coefficient $D$, pressure $P$ \cite{PRESSURE}
and potential energy $U$.  We calculate $D$ using the Einstein relation
$D=\frac{1}{2d}\lim_{t \rightarrow \infty}\frac{<\Delta r(t)^2>}{t}$,
where $d$ is the system dimension.  After a short initial period of
increase, we observe that $D$ and $U$ decrease with time and reach
plateaus [Fig.~\ref{plateau}].  The diffusion coefficient approaches
zero, which is consistent with the possible formation of a quasicrystal
phase.  The isointensity peaks also sharpen with the duration of
annealing.  Due to MD time constraints, we are not sure that we reach
the potential energy of a perfect quasicrystal, which is expected to be
comparable to the energies, $U_{pI}=-4\frac{2}{5}$ and
$U_{pII}=-4\frac{1}{3}$, of the pentagonal crystals.  The lowest
potential energy reached is $U_{qc}=-4.25$.

We observe the spontaneous formation of the quasicrystal phase in the
range of temperatures between $T=0.190$ and
$T=0.205$.  As we heat either the amorphous phase or the
quasicrystal above $T=0.21$ , the square crystal forms,
consistent with the Helmholtz potential estimations of Fig.~\ref{fe}.

Next we address the question of whether the quasicrystal phase is
stable, by comparing the values of the Helmholtz potential for the three
crystal types. As can be seen [Fig.~\ref{diff-fig}c,d], the structure of
the quasicrystal arises from the bending rows of pentagons which locally
resemble the pentagonal crystals of either type I or II.  We assume that
local arrangements of particles corresponding to a square crystal are
defects \cite{defects} that would be absent in the perfect quasicrystal.
If we assume that the local arrangement of the quasicrystal is similar
to a combination of the local arrangements in the pentagonal crystal I
and the pentagonal crystal II, we can approximate the Helmholtz
potential of the quasicrystal by the average Helmholtz potential of the
two pentagonal crystals.  Because the quasicrystals have a positive
entropy contribution to the total entropy due to their degeneracy
\cite{WOJ}, we add an additional term $-TS_{c}$ to the original estimate
of the Helmholtz potential energy. Here $S_{c}$ is the entropy due to
the possible configurations of the quasicrystal.

\subsection{Entropy}

We estimate $S_{c}$ as the logarithm of the number of configurations
formed by $n$ pentagons in the quasicrystal.  A single pentagon can be
oriented in two possible ways when attached side by side to an existing
row of pentagons. Neglecting the interaction between adjacent rows, we
can estimate the upper bound for the number of configurations as
$2^{n}$, where $n$ is the total number of pentagons in the
quasicrystal. Note that at point A on Fig.~\ref{fe}, the Helmholtz
potentials of both pentagonal crystals coincide, so an additional
$-TS_{c}$ term should stabilize the quasicrystal in the vicinity of
point A.

To better estimate $S_c$, we notice that the bending rows of pentagons
forming the quasicrystal resemble a compact self-avoiding random walk on
the hexagonal lattice. The number of such walks grows as $Z^{n}$ where
$Z\approx1.3$ and $n$ is the number of steps \cite{SERGEY}. Since the
formation of one pentagon in the midst of a perfect square crystal
lowers the energy of the system by $U=-1 $, we estimate $n$ to be
$(U_{qc}-U_{sq})N$.  Assuming that the ground state energy of the
quasicrystal is between $U_{pI}$ and $U_{pII}$, the number of pentagons
in the quasicrystal, should not be smaller than the number of pentagons
in the crystal of type II (which is the pentagonal crystal with the
lesser number of pentagons and has $n=\frac{1}{3}N$). We estimate the
entropy of configuration per particle to be $S_c \approx\ln(Z^{n})/N=
\frac{1}{3}\ln(1.3)=0.087$.  Thus, the quasicrystal should be more
stable than the pentagonal crystals between $T=0.16$ and $T=0.23$ ,
where the gap between the Helmholtz potential of the pentagonal crystals
is smaller than the configuration term $TS_{c}$ which ranges from
$0.014$ to $0.020$ in the interval where $T$ increases from $0.16$ to
$0.23$. Since the $TS_{c}$ term lowers the Helmholtz potential of the
obtained quasicrystal configuration below the Helmholtz potentials of
the two pentagonal crystals, it is likely that the obtained state with
five-fold rotational symmetry is not the coexistence of type I and II
pentagonal crystals, but is a stable quasicrystalline phase.  A more
rigorous investigation of this problem would either require the
construction of a perfect Penrose tiling \cite{SEN,CLH} or of a random
tiling \cite{QUAN,RICH} involving the local structures of crystals type
I and II.

\section{Discussion}

To summarize, perfect pentagonal crystals of type I and II do not form
spontaneously during the time scales of our study.  Instead, the
quasicrystal, having long-range, five-fold orientational order with no
translational order, forms from the coalescence of pentagons present in
the starting amorphous phase.  The starting amorphous configuration must
initially be quenched at a low enough temperature in order to prevent
crystallization to the square phase.  Moreover, the amorphous phase must
be carefully thermalized at the quench temperature, as we have observed
that, upon heating a poorly equilibrated amorphous phase with a higher
concentration of squares, the system phase separates into regions of
pentagons and squares. If the starting amorphous phase does not have a
sufficient concentration of pentagons, the quasicrystal will not form:
large regions of square crystal will inhibit the long range
order of pentagons and thus not give rise to the 10 diffraction peaks in
the isointensity plot.  It is interesting to notice that the bending
rows observed in our quasicrystal could resemble the stripe structure of
a spinodal decomposition \cite{LUBE}.  Anyhow, in the case of spinodal
decomposition, the diffraction pattern would be similar to that of an
amorphous structure.

Before concluding, we note that Jagla, using Monte Carlo simulations,
recently reported the existence of quasicrystals in a two-dimensional,
monodisperse system of hard spheres interacting with a {\it purely
repulsive potential} \cite{JAGLA}.  The quasicrystal we observe has a
different structure from that modeled by Jagla: our quasicrystal is not
a ground state structure and forms only at nonzero temperature.  Also,
formation of quasicrystals in monodisperse systems has been observed
using complex radially symmetric potentials both in two dimensions
\cite{QUAN} and three dimensions \cite{DZUG,DENT}.  To the best of our
knowledge, the quasicrystal found in our simulations has a structure
different from those previously studied.

We are very grateful to the late Shlomo Alexander, who pointed out the
possibility of the formation of quasicrystals in the square-well
potential, and we dedicate this work to his memory. We thank R. Hurt and
his colleagues at Brown University for encouraging this project in its
early stages, L. A. N. Amaral, C. A. Angell, E. Jagla, J. E. McGarrah,
C. J. Roberts, R. Sadr, F. Sciortino, F. W. Starr, A. Umansky, Masako
Yamada for helpful interactions, and the referee for constructive
criticism. We also thank DOE and NSF for financial support.

\pagebreak

\pagebreak

\begin{table}
\caption{Energy $U$, entropy $S$ and the Helmholtz potential $A$ at
temperature $T=0.2$ where the quasicrystal is found}
\medskip
\begin{tabular}{c||cc|cc|cc}
\tableline
Crystal  & $U$ && $S$  && $A (T=0.2)$ & \\ \hline
Pentagonal I & $-4\frac{2}{5}$ && $1.259\pm0.028$  && $-4.652$& \\ \hline
Pentagonal II& $-4\frac{1}{3}$ && $1.603\pm0.0052$  && $-4.654$&\\ \hline
Square       & $-4$ && $3.247\pm0.021$ && $-4.649$& \\ \end{tabular}
\label{simulation-table}
\end{table}

\newbox\figa 
\setbox\figa=\psfig{figure=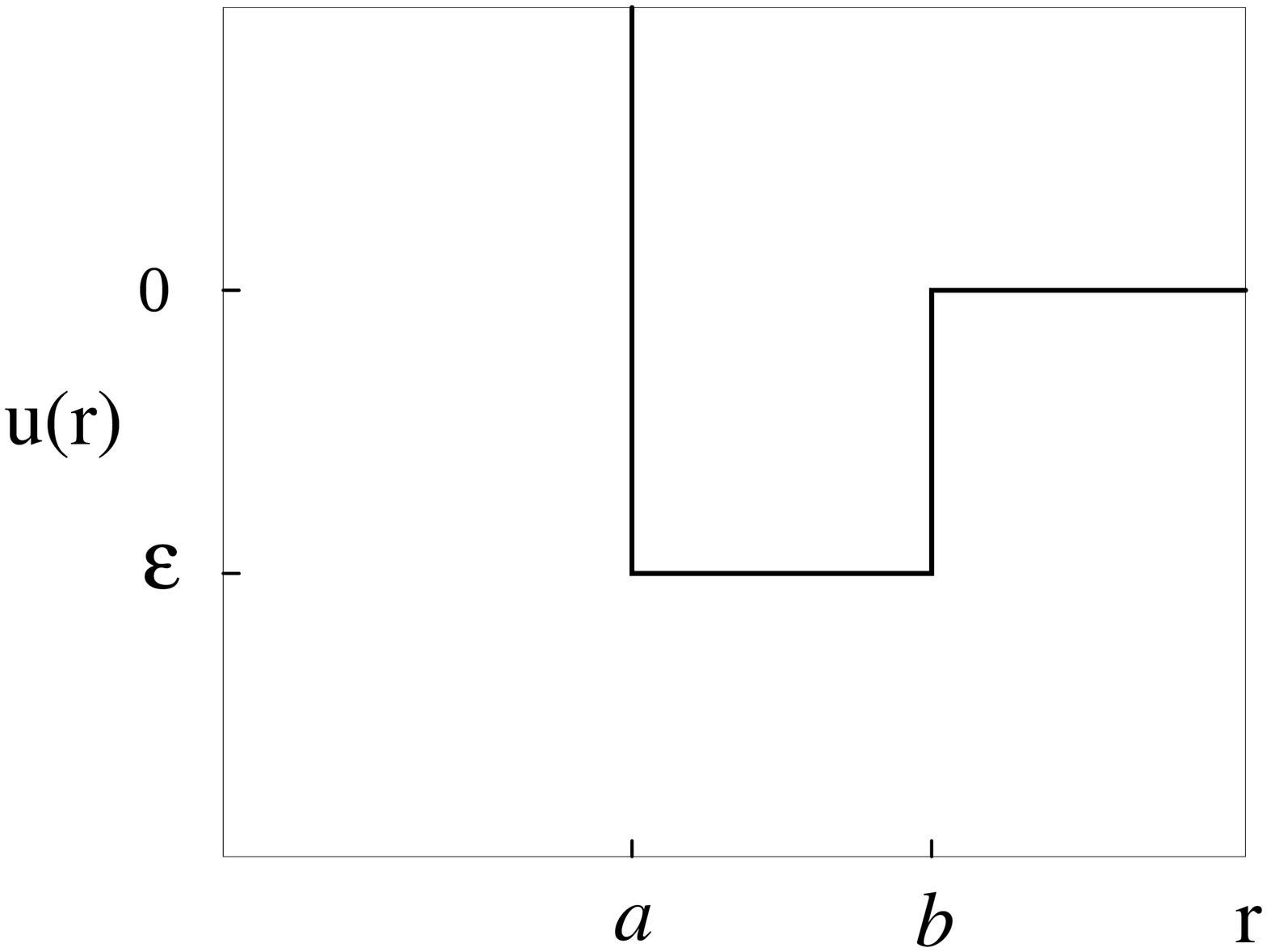,width=10cm,angle=-90}
\begin{figure*}[htbp]
\begin{center}
\leavevmode
\centerline{\box\figa}
\narrowtext
\caption{The square-well potential. The ratio of the attractive distance $b$
to the hard core repulsive distance $a$ is $b/a=\protect\sqrt{3}$. The
depth of the square-well $\epsilon=-1.0$ is the interaction energy per
pair of particles.}
\label{sw_pot} 
\end{center} 
\end{figure*}

\newbox\figa 
\setbox\figa=\psfig{figure=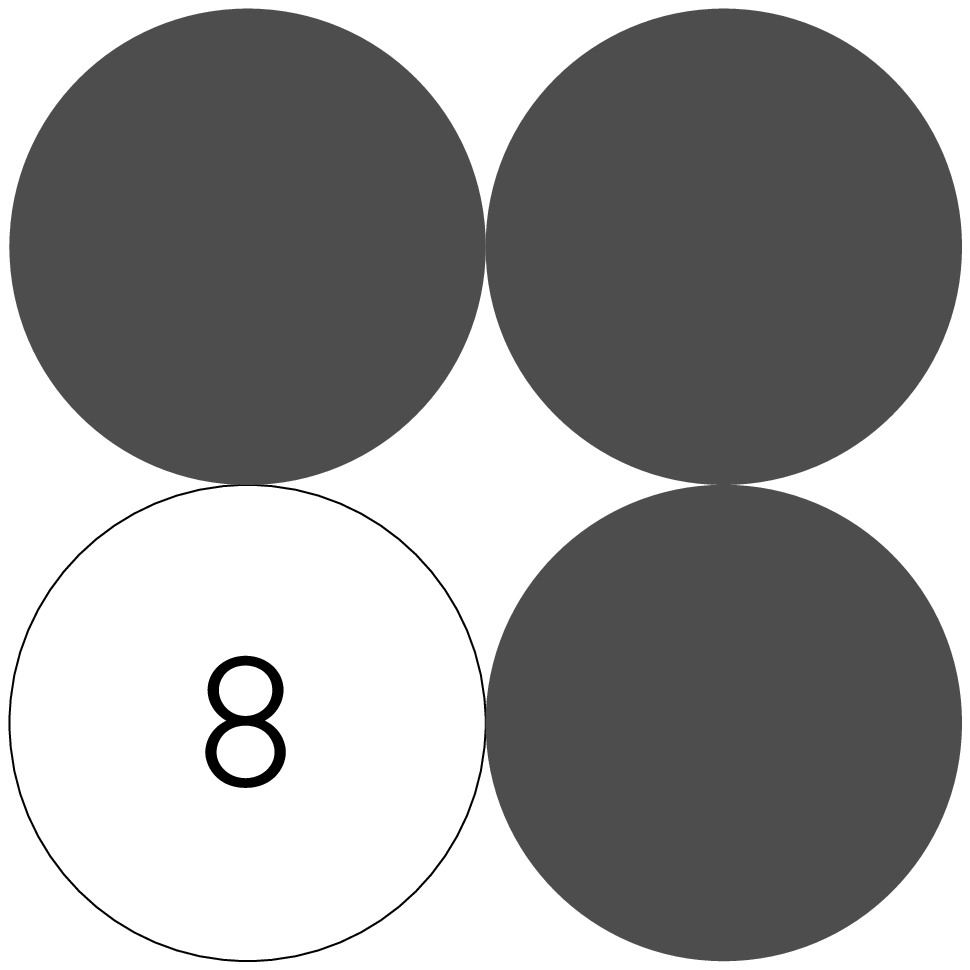,width=10cm,angle=0}
\begin{figure*}[htbp]
\begin{center}
\leavevmode
\centerline{\box\figa}
\end{center}
\end{figure*}

\newbox\figa 
\setbox\figa=\psfig{figure=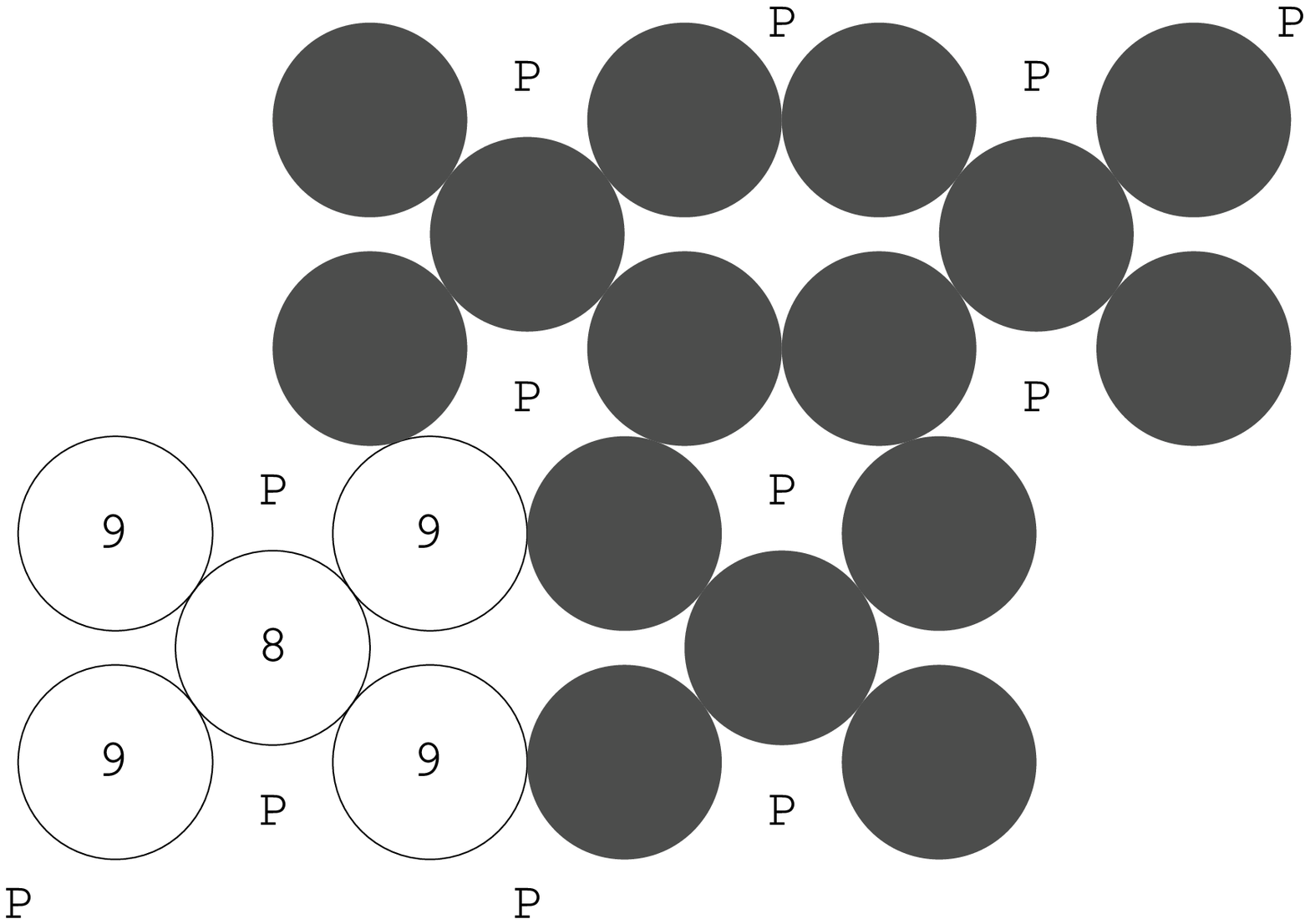,width=10cm,angle=0}
\begin{figure*}[htbp]
\begin{center}
\leavevmode
\centerline{\box\figa}
\end{center}
\end{figure*}

\newbox\figa 
\setbox\figa=\psfig{figure=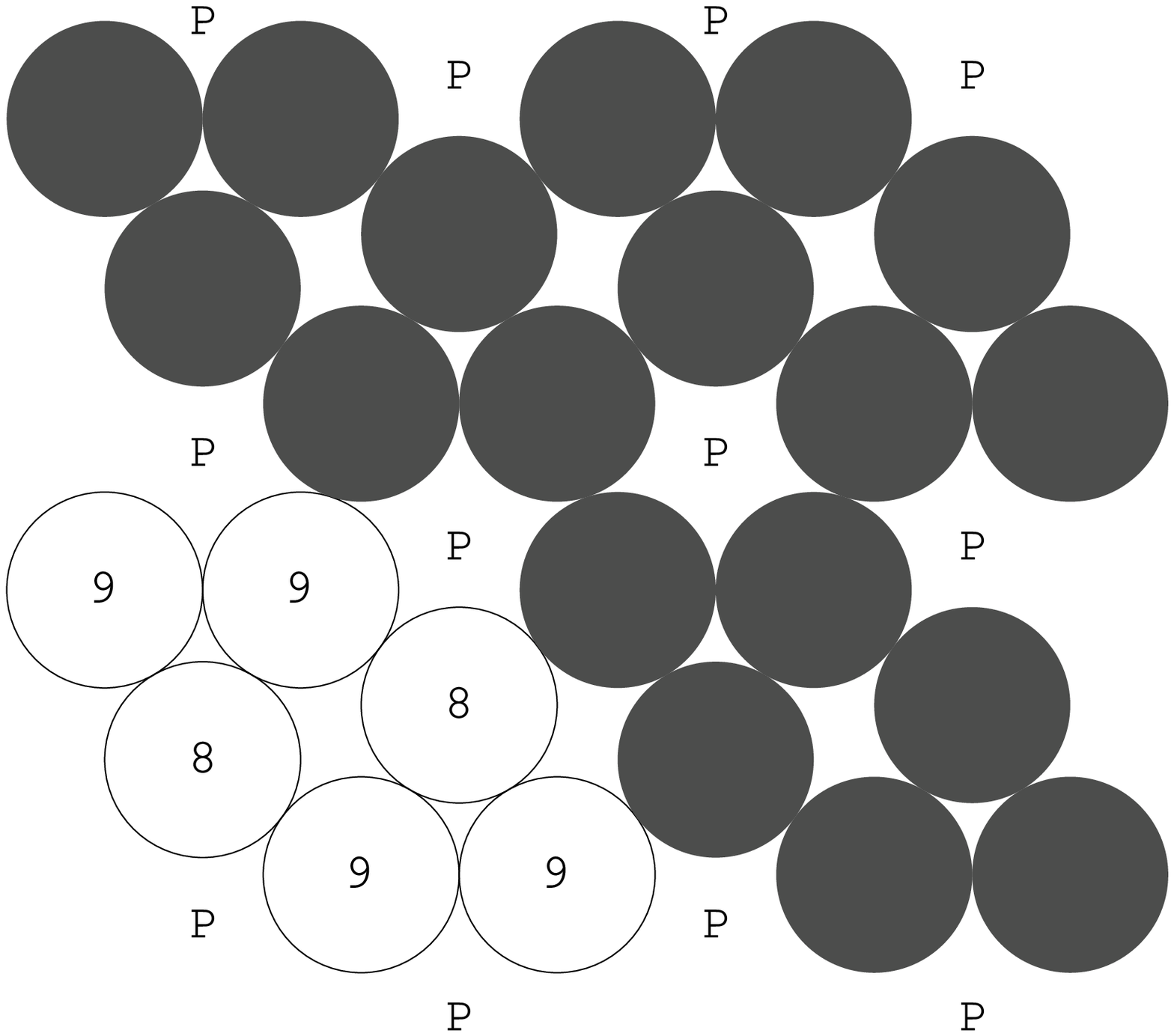,width=10cm,angle=0}
\begin{figure*}[htbp]
\begin{center}
\leavevmode
\centerline{\box\figa}
\narrowtext
\caption{Repeating segments of the three crystals. (a) In the square
crystal, each particle interacts with $8$ nearest neighbors. The square
crystal is constituted by particles interacting with a square
geometry.\  (b) In type I pentagonal crystals, $\frac{1}{5}$ of
the particles have $8$ neighbors and $\frac{4}{5}$ of the particles have
$9$ neighbors. (c) In type II pentagonal crystals, $\frac{1}{3}$ of the
particles have $8$ neighbors and $\frac{2}{3}$ of the particles have $9$
neighbors. Five particle pentagons, denoted by letter ``P'', form the
pentagonal crystals.  The particles indicated in white are the particles
in a basic cell that can be used to construct the crystal by
translation; there are respectively one, five and six particles in the
unit cell of the square, pentagonal I and pentagonal II crystals. }
\label{3xtals}
\end{center}
\end{figure*}

\newbox\figa 
\setbox\figa=\psfig{figure=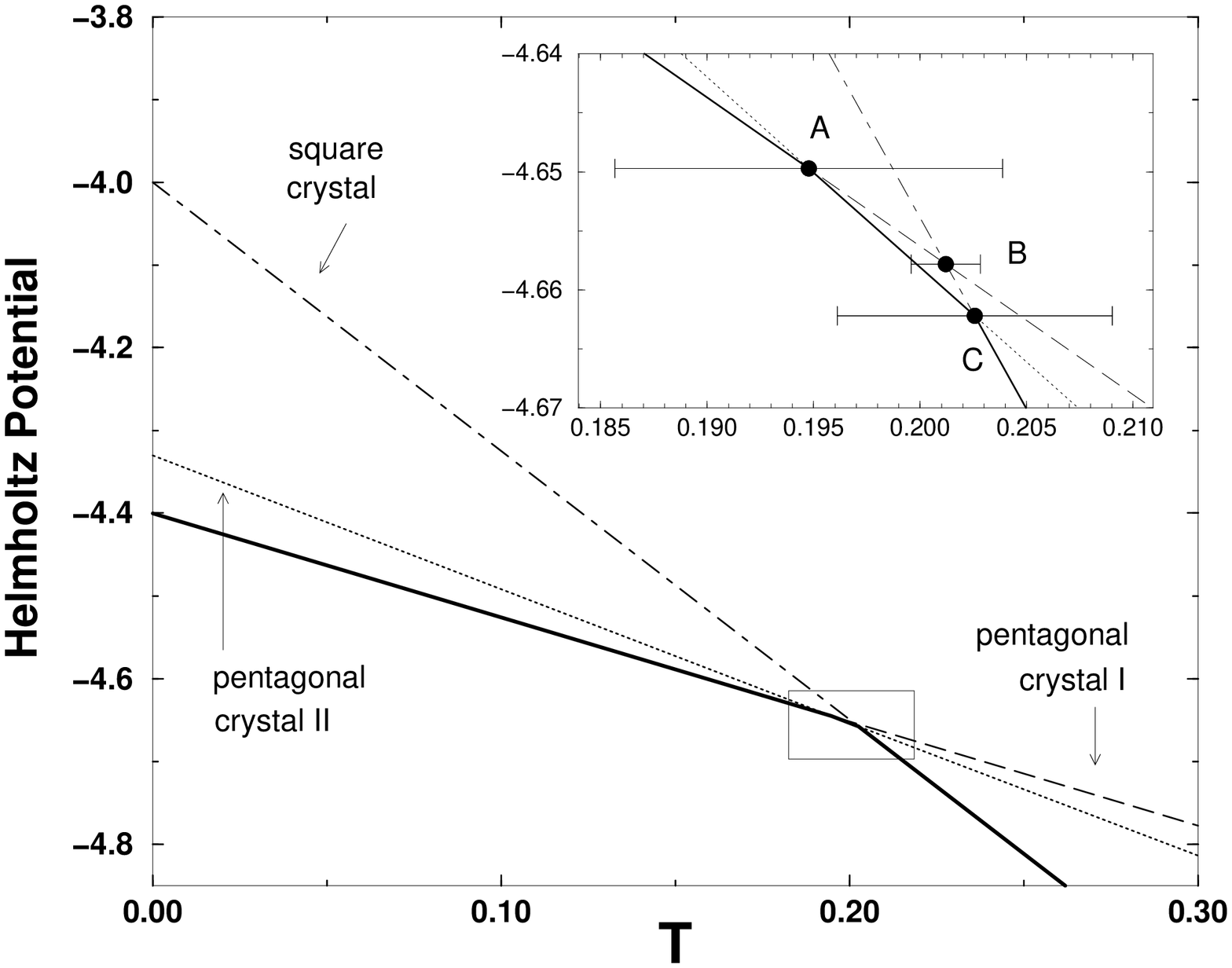,width=10cm,angle=-90}
\begin{figure*}[htbp]
\begin{center}
\leavevmode
\centerline{\box\figa}
\narrowtext
\caption{The Helmholtz potentials of pentagonal crystals of type I and
II and the square crystal at various temperatures. Points A, B, and C,
of the inset, indicate the intersections of the Helmholtz potential
lines at $T_{A}=0.195\pm0.010$,
$T_{B}=0.201\pm0.005$,
$T_{C}=0.203\pm0.006$.  The solid line indicates the
lowest Helmholtz potential: below $T_{A}$ the type I pentagonal crystal
is the most stable, between $T_{A}$ and $T_{C}$ the type II pentagonal
crystal is the most stable, and above $T_{C}$ the square crystal is the
most stable. }
\label{fe}
\end{center}
\end{figure*}

\newbox\figa 
\setbox\figa=\psfig{figure=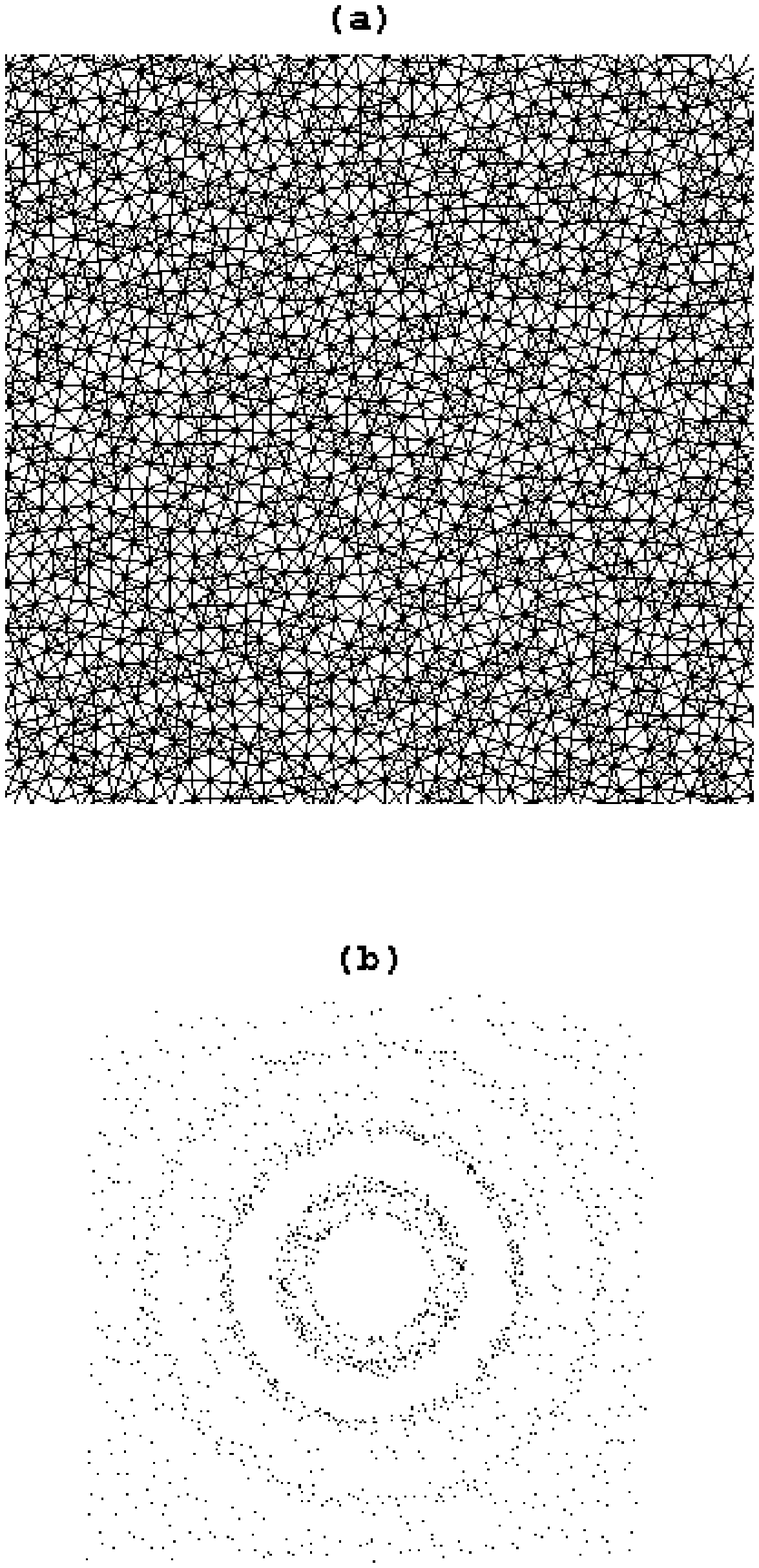,width=10cm,angle=0}
\begin{figure*}[htbp]
\begin{center}
\leavevmode
\centerline{\box\figa}
\end{center}
\end{figure*}

\newbox\figa 
\setbox\figa=\psfig{figure=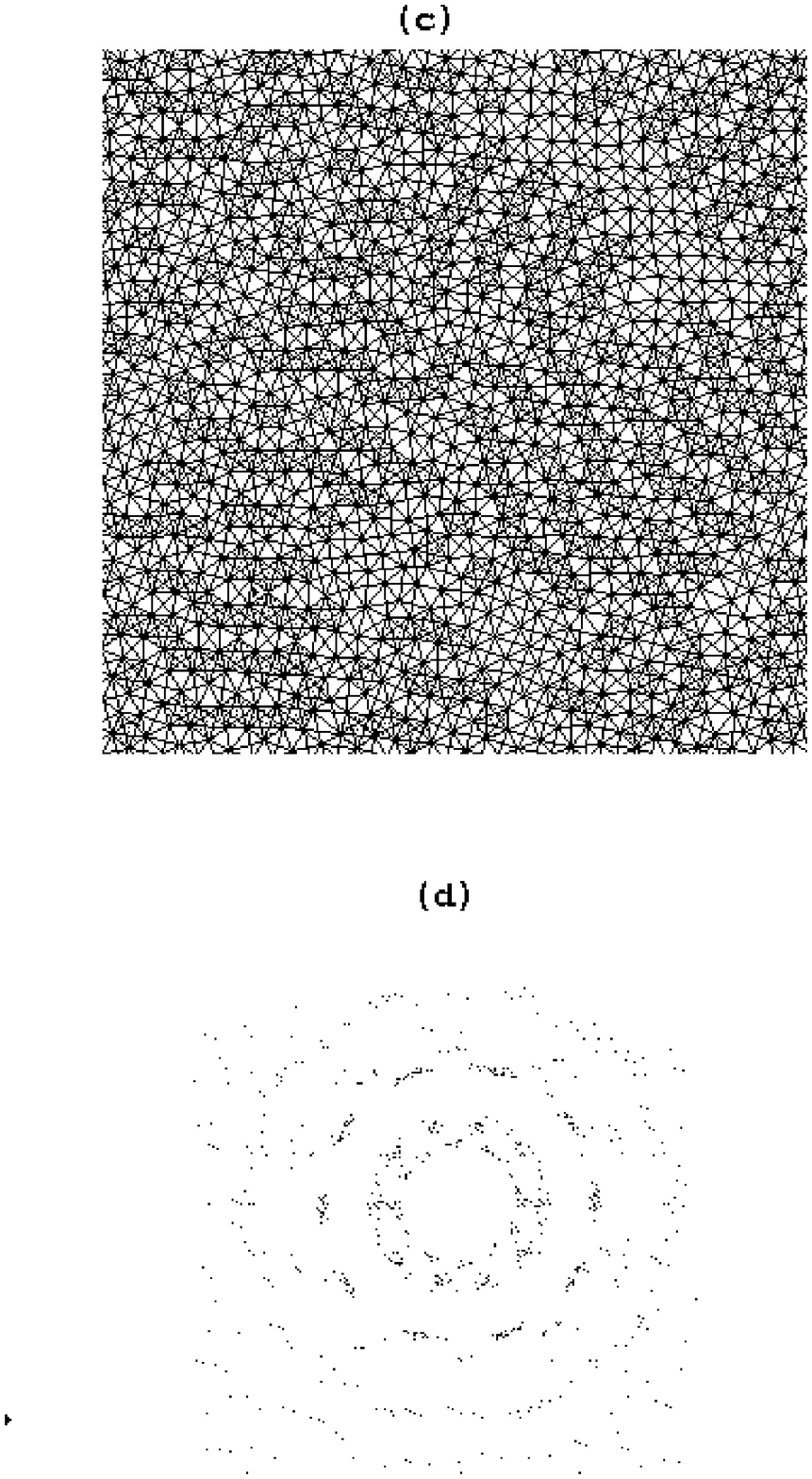,width=10cm,angle=0}
\begin{figure*}[htbp]
\begin{center}
\leavevmode
\centerline{\box\figa}
\narrowtext
\caption{Amorphous and quasicrystal phases are shown along with their
corresponding isointensity plots: the simulated equivalent to a
crystallographic diffraction pattern, given by the Fourier transform of
the density function: the darkness is proportional to the amplitude of
the Fourier transform. On the original system snapshots (a,c), pentagons
are indicated by the shaded areas and lines indicate interacting pairs
of particles.  (a) Uniformly distributed pentagons in the amorphous
phase give rise to the (b) homogeneous rings in the isointensity plot.
(c) The pentagons in the quasicrystal phase have coalesced in curved
rows that run approximately parallel to one another, in contrast to part
(a) where the rows are much less apparent and are not even approximately
parallel.  (d) The ten isointensity peaks of the quasicrystal.  }
\label{diff-fig} 
\end{center} 
\end{figure*}

\newbox\figa 
\setbox\figa=\psfig{figure=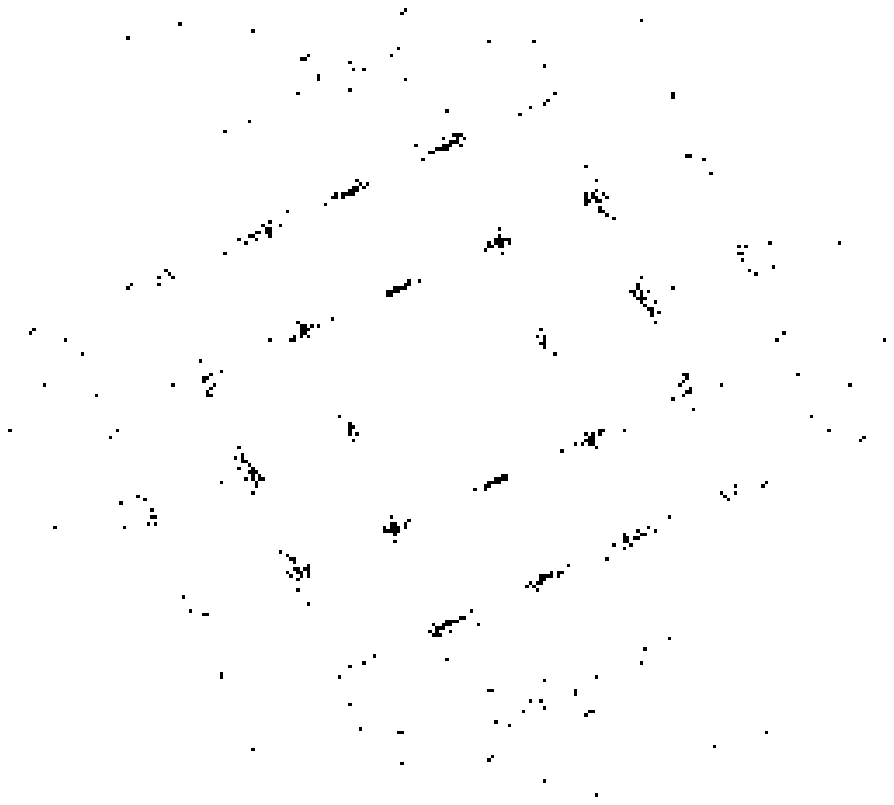,width=10cm,angle=-90}
\begin{figure*}[htbp]
\begin{center}
\leavevmode
\centerline{\box\figa}
\end{center}
\end{figure*}

\newbox\figa 
\setbox\figa=\psfig{figure=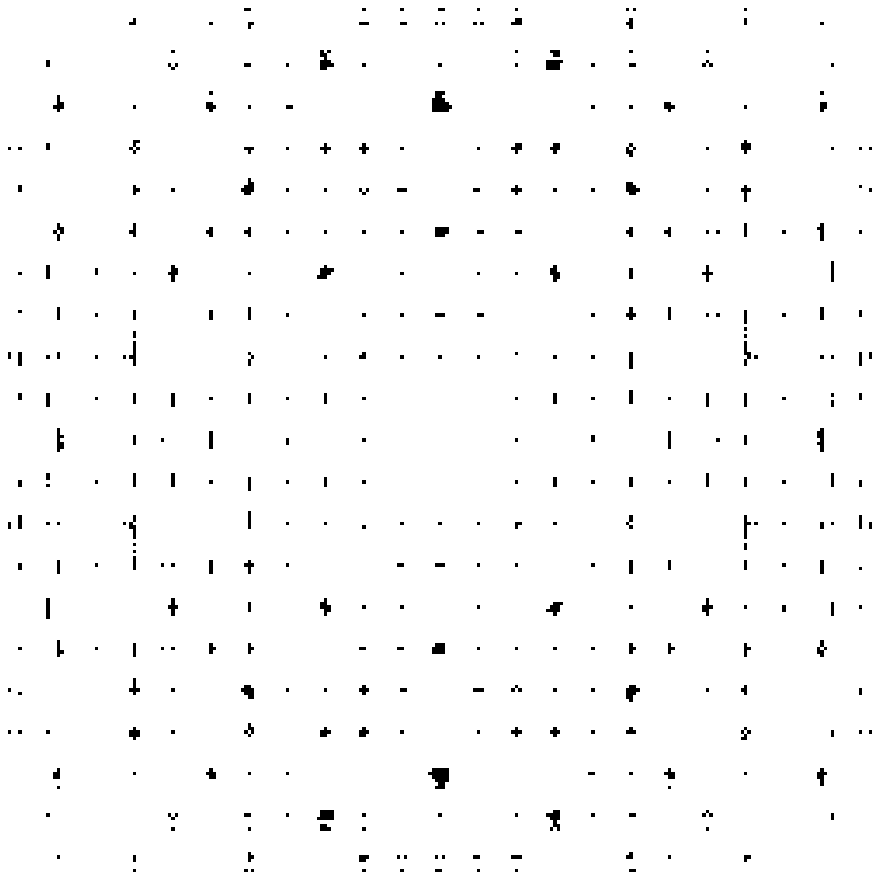,width=10cm,angle=-90}
\begin{figure*}[htbp]
\begin{center}
\leavevmode
\centerline{\box\figa}
\end{center}
\end{figure*}

\newbox\figa 
\setbox\figa=\psfig{figure=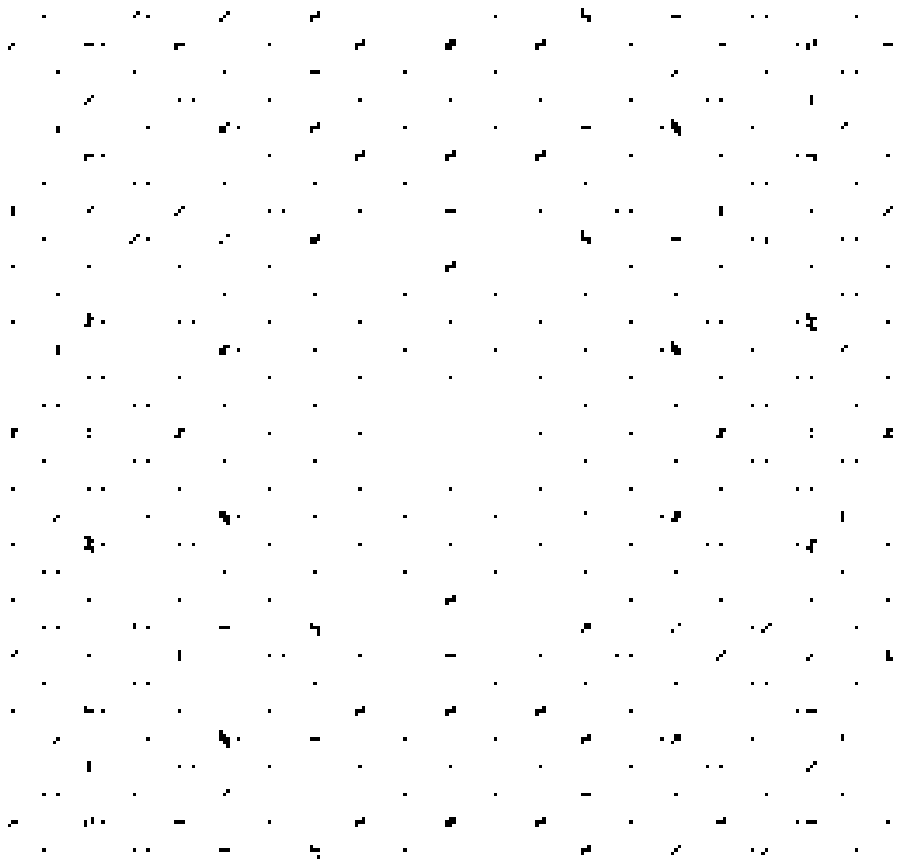,width=10cm,angle=-90}
\begin{figure*}[htbp]
\begin{center}
\leavevmode
\centerline{\box\figa}
\narrowtext
\caption{The (a) square crystal, (b) type I pentagonal crystal and (c) type II
pentagonal crystal isointensity plots.}
\label{pxtals-fig}
\end{center}
\end{figure*}

\newbox\figa 
\setbox\figa=\psfig{figure=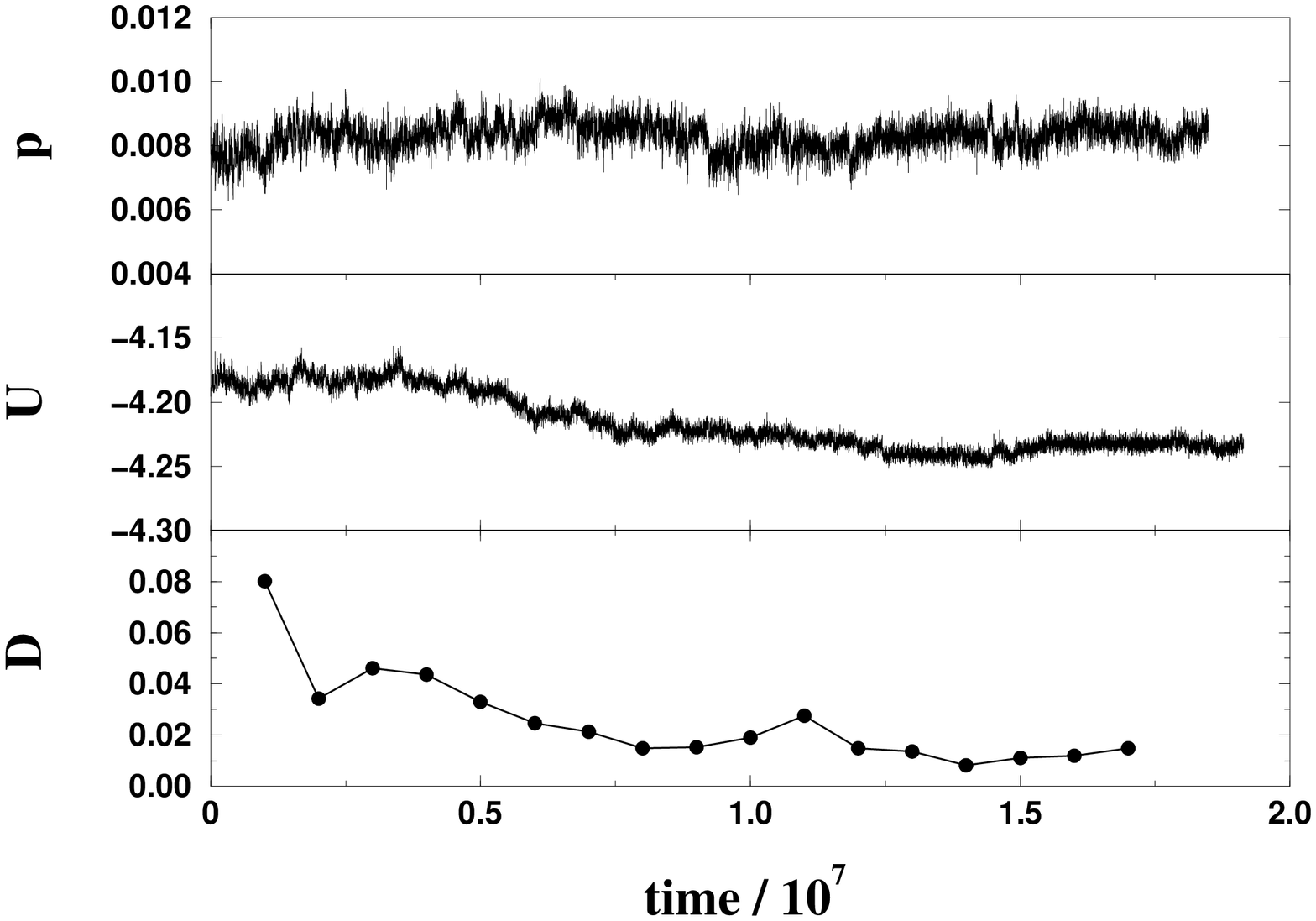,width=10cm,angle=-90}
\begin{figure*}[htbp]
\begin{center}
\leavevmode
\centerline{\box\figa}
\narrowtext
\caption{Behavior of pressure, $P$, potential energy per particle, $U$ and
diffusion coefficient, $D$ versus time when the system, initially in the
amorphous phase, is equilibrated at $T=0.205$. The density is
$\rho=0.857$ and the number of particles is $N=961$}
\label{plateau}
\end{center}
\end{figure*}


\begin{references}
\bibitem{FN529}To verify the existence of the quasicrystal, we simulated
the same density in a smaller system of 529 particles.  At
$T=0.2$, the 10 five-fold diffraction peaks appeared,
giving evidence to the formation of the quasicrystalline phase.
\bibitem{TIME} Time units are defined so that temperature,
$T=<mv^{2}/2>$. For our parameters of N, V, and T=0.1, we find that the
average collision interval is approximately one time unit.
\bibitem{qrq}To form the amorphous phase from the high temperature
liquid, a quench rate of $\kappa=1$ is used. The amorphous phase is
annealed at a slower quench rate of $\kappa=0.01$.
\bibitem{PRESSURE}For $\rho=0.857$, pressure at this density is almost
constant, as energy continually decreases (Fig.~\ref{plateau}). For
densities slightly below or above $\rho = 0.857$, pressure suffers a
constant drift during the time scales accessible to our simulations,
indicating that the system, at these densities, is further from
equilibrium than when $\rho=0.857$.
\end{references}
\end{document}